\documentclass[a4paper,11pt]{article}
\usepackage{pos}
\usepackage{dirtytalk}
\usepackage{caption}
\usepackage{subcaption}
\usepackage{wrapfig}
\usepackage{tikz}
\usepackage{mathtools}

\newcommand{\Z}{{\mathbb{Z}}}
\newcommand{\R}{{\mathbb{R}}}

\newcommand{\presc}[1]{\prescript{C}{}{#1}}
\newcommand{\presg}[1]{\prescript{G}{}{#1}}

\newcommand{\SU}{\mathrm{SU}}
\newcommand{\U}{\mathrm{U}}

\newcommand{\pqty}[1]{\left( #1 \right)}
\newcommand{\bqty}[1]{\left[ #1 \right]}
\newcommand{\abs}[1]{\left\lvert #1\right\rvert}

\newcommand{\expval}[1]{\langle #1 \rangle}

\newcommand {\dv}[3][ ]{
  \ifx #1 { }
    \frac{d #2}{d #3}
  \else
    \frac{d^{#1} #2}{d #3^{#1}}
  \fi
}
\newcommand {\pdv}[3][ ]{
  \ifx #1 { }
    \frac{\partial #2}{\partial #3}
  \else
    \frac{\partial^{#1} #2}{\partial #3^{#1}}
  \fi
}
\newcommand {\fdv}[3][ ]{
  \ifx #1 { }
    \frac{\delta #2}{\delta #3}
  \else
    \frac{\delta^{#1} #2}{\delta #3^{#1}}
  \fi
}

\renewcommand{\Im}{\operatorname{Im}}
\renewcommand{\Re}{\operatorname{Re}}

\title{Confined Charged Particles in C-periodic Volumes}

\author[a]{G. Kanwar}
\author[a]{A. Mariani}
\author[b]{J. C. Pinto Barros}
\author*[a]{U.-J. Wiese}

\affiliation[a]{Albert Einstein Center for Fundamental Physics,\\ Institute for Theoretical Physics, Bern University\\
Sidlerstrasse 5, CH-3012 Bern, Switzerland}

\affiliation[b]{Institut für Theoretische Physik, ETH Zürich,\\
Wolfgang-Pauli-Str. 27, 8093 Zürich, Switzerland}

\emailAdd{kanwar@itp.unibe.ch}
\emailAdd{mariani@itp.unibe.ch}
\emailAdd{jpinto@phys.ethz.ch}
\emailAdd{wiese@itp.unibe.ch}

\abstract{Charged particles in an Abelian Coulomb phase are non-local infraparticles that are surrounded by a cloud of soft photons which extends to infinity. Gauss’ law prevents the existence of charged particles in a periodic volume. In a $C$-periodic volume, which is periodic up to charge conjugation, on the other hand, charged particles can exist. This includes vortices in the $3$-d XY-model, magnetic monopoles in $4$-d $\U(1)$ gauge theory, as well as protons and other charged particles in QCD coupled to QED. In four dimensions non-Abelian charges are confined. Hence, in an infinite volume non-Abelian infraparticles cost an infinite amount of energy. However, in a $C$-periodic volume non-Abelian infraparticles (whose energy increases linearly with the box size) can indeed exist. Investigating these states holds the promise of deepening our understanding of confinement.}

\FullConference{%
  The 39th International Symposium on Lattice Field Theory, LATTICE2022\\ 8th-13th August, 2022\\
  University of Bonn, Bonn, Germany
}

\begin{document}

\maketitle

\section{Introduction: C-periodic Boundary Conditions}
The most common boundary conditions for quantum field theories in a finite volume are periodic boundary conditions. These have several advantages, for example they preserve translational invariance and they are simple to implement in numerical simulations. However, they also have some drawbacks. In particular, the topology of space-time induced by periodic boundaries in $d$-dimensions is that of the $d$-dimensional torus $T^d$. Since the torus is a manifold without boundary, Gauss' law implies that the overall charge on the torus is always zero,
\begin{equation}
    Q = \int_{T^d} d^d\,x\, \vec{\nabla}\cdot\vec{E} = \int_{\partial T^d} d\vec{S} \cdot \vec{E} = 0 \ .
\end{equation}
Therefore periodic boundary conditions are unsuitable whenever an overall net charge is required to be present in the system. The most general boundary conditions which preserve translational invariance are those where one performs a symmetry transformation when crossing a boundary. Among this class of boundary conditions which preserve translational invariance are $C$-periodic boundary conditions \cite{PolleyWiese, KronfeldWiese}, where one performs a charge conjugation twist at the spatial boundary,
\begin{align}\label{eq:charge conjugation definition}
    A_\mu(\vec{x} + L_i \vec{e}_i, t) &= \presc{A}_\mu(\vec{x}, t) = - A_\mu(\vec{x}, t) \ , \nonumber\\
    \Phi(\vec{x} + L_i \vec{e}_i, t) &= \presc{\Phi}(\vec{x}, t) = \Phi(\vec{x}, t)^* \ , \nonumber\\
    G_\mu(\vec{x} + L_i \vec{e}_i, t) &= \presc{G_\mu}(\vec{x}, t) = G_\mu(\vec{x}, t)^* \ , \\
    \psi(\vec{x} + L_i \vec{e}_i, t) &= \presc{\psi}(\vec{x}, t) = C \bar{\psi}(\vec{x}, t)^T \ , \nonumber
\end{align}
for Abelian gauge fields, scalar fields, non-Abelian gauge fields and fermion fields respectively, where $L_i$ is the extent of the finite volume in spatial direction $\vec{e}_i$. Note that eq.~\eqref{eq:charge conjugation definition} holds for each spatial direction separately, i.e.\ no summation over $i$ is implied.

Imposing $C$-periodic boundary conditions affects the symmetries of the theory. As previously remarked, translational invariance is preserved but it mixes with charge conjugation. Therefore momentum quantization is now determined by the $C$-parity of a state, i.e.
\begin{align}
    &C = +: \quad \Re\Phi(\vec{x} + L_i \vec{e}_i, t) = \Re \Phi(\vec{x}, t) \quad \quad\quad \implies \quad \quad p_i = \frac{2\pi n_i}{L_i} \ ,\nonumber\\
    &C = -: \quad \Im\Phi(\vec{x} + L_i \vec{e}_i, t) = -\Im \Phi(\vec{x}, t) \quad \quad \implies \quad \quad p_i = \frac{2\pi (n_i+\tfrac12)}{L_i} \ .
\end{align}
$C$-periodic boundary conditions also affect gauge transformations, baryon number, and the center symmetry \cite{WieseCG}. In particular, gauge transformations are required to be $C$-periodic,
\begin{align}\label{eq:c periodic gauge transformations}
    &A_\mu'(x) = A_\mu(x)-\partial_\mu \alpha(x) \quad \quad\quad\quad \quad\quad \implies \quad \quad \alpha(\vec{x} + L_i \vec{e}_i, t) = -\alpha (\vec{x}, t) \ ,\nonumber\\
    &G_\mu'(x) = \Omega(x) (G_\mu(x)+\partial_\mu) \Omega(x)^\dagger \quad \quad\quad \implies \quad \quad \Omega(\vec{x} + L_i \vec{e}_i, t) = \Omega(\vec{x}, t)^* \ ,
\end{align}
in the Abelian and non-Abelian cases, respectively. Since in a $C$-periodic volume quarks and anti-quarks can mix, baryon number is partially broken to $\Z(2)$ fermion number, $\U(1)_B \to \Z(2)_F$. The $\Z(N)$ center symmetry for $\SU(N)$ gauge theories is also partially or completely broken by $C$-periodic boundary conditions depending on $N$. For $z \in \Z(N) \subset \SU(N)$ a center transformation $\Omega(\vec{x}, t+\beta) = z\Omega(\vec{x},t)$ (see eq.~\eqref{eq:c periodic gauge transformations}), where $\beta$ is the time extent of the finite volume, must satisfy
\begin{equation}
    \Omega(\vec{x} + L_i \vec{e}_i, t+\beta)^* = \Omega(\vec{x}, t+\beta) =z \Omega(\vec{x}, t) = z \Omega(\vec{x}+L_i \vec{e}_i, t)^* = z^2 \Omega(\vec{x} + L_i \vec{e}_i, t+\beta)^* \ .
\end{equation}
Therefore $z^2=1$, or equivalently $z=z^*$. This means that for even $N$ the center symmetry is broken to a subgroup, $\Z(N) \to \Z(2)$, while for odd $N$ it is completely broken, $\Z(N) \to 1$.

It is also interesting to construct $G$-periodic boundary conditions, where one performs a $G$-parity transformation at the spatial boundary \cite{WieseCG}. Since $G$-parity is a combination of charge conjugation with a flavor transformation, $G$-periodic boundary conditions reduce to $C$-periodic boundary conditions for flavor singlets such as gluons. For the quarks, in the chiral basis $G$-periodic boundary conditions are given by
\begin{align}
    \psi_R(\vec{x} + L_i \vec{e}_i, t) &= \presg{\psi}_R(\vec{x}, t) = \sigma_2 \otimes \sigma_2 \, \bar{\psi}_L(\vec{x}, t)^T \ ,\nonumber\\
    \psi_L(\vec{x} + L_i \vec{e}_i, t) &= \presg{\psi}_L(\vec{x}, t) = -\sigma_2 \otimes \sigma_2 \, \bar{\psi}_R(\vec{x}, t)^T \ .
\end{align}
Interestingly, $G$-periodicity breaks the chiral symmetry $\SU(2)_R \times \SU(2)_L \to \SU(2)_{L=R}$ much like a fermion mass term.

In these proceedings, we explore the use of $C$-periodic boundary conditions in the context of lattice field theories. In Section \ref{sec:vortex} we show that topological excitations in the $(2+1)$-d $\U(1)$ scalar field theory can be mapped to charged particles in a dual theory which can be studied via $C$-periodic boundary conditions. This allows a fully non-perturbative construction of these topological excitations. In Section \ref{sec:quarks} we show that $C$-periodic boundaries allow the construction of a single-quark state in a finite volume for $\SU(N)$ gauge theories with $N \geq 4$ and $N$ even, thus potentially deepening our understanding of confinement.

\section{Vortex in the (2+1)-d U(1) Scalar Field Theory}\label{sec:vortex}

\begin{figure}
    \centering
    \includegraphics{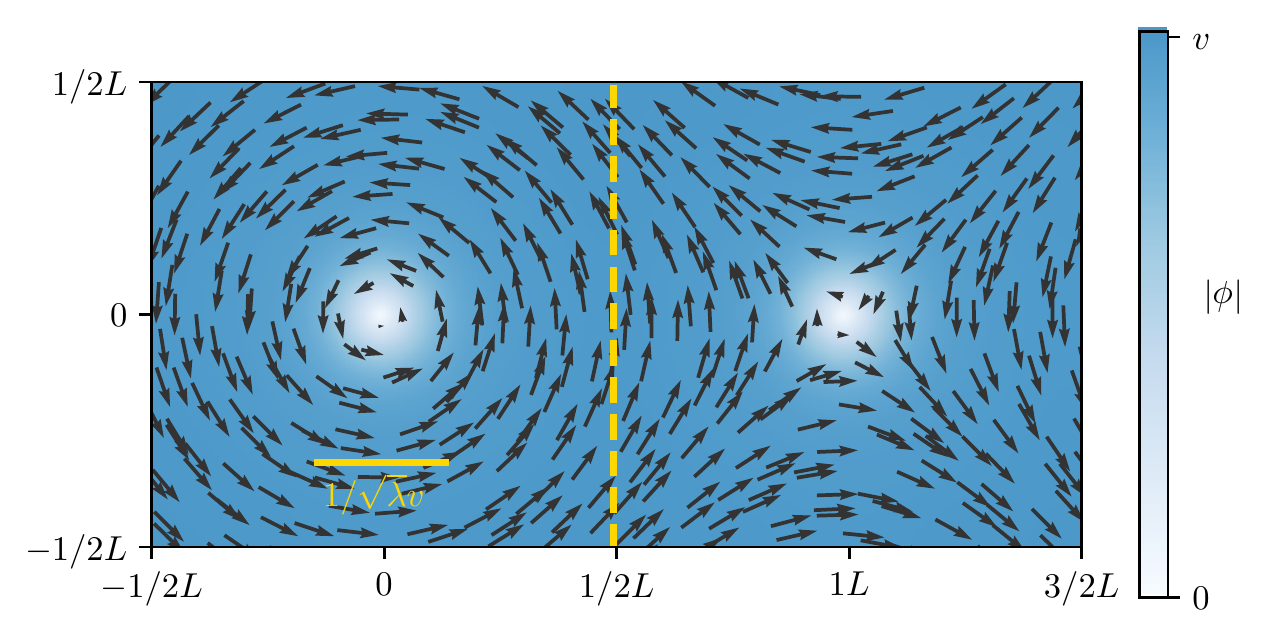}
    \caption{Classical static vortex solution of the $(2+1)$-d $\U(1)$ scalar field theory with twisted $C$-periodic boundary conditions, taken from \cite{VortexMass}. The arrows represent the complex field $\Phi(\vec{x})$. The right side of the figure is the $C$-periodic copy of the left side.}
    \label{fig:classical vortex}
\end{figure}

The $(2+1)$-d $\U(1)$ scalar field theory is described by the classical Lagrangian
\begin{equation}
    \mathcal{L} = \frac12 \partial_\mu \Phi^* \partial^\mu \Phi - \frac{\lambda}{4!} \pqty{\abs{\Phi}^2 - \upsilon^2}^2 \ ,
\end{equation}
where $\Phi$ is a complex scalar field. 
Imposing spatial $C$-periodic boundary conditions on the complex scalar field $\Phi$, the theory admits classical static vortex solutions of the form $\Phi(\vec x) = f(r) e^{i\varphi}$ in the broken phase, where $\vec{x} = (r \cos{\varphi}, r \sin{\varphi})$. The numerical solution of the resulting equations of motion is shown in Fig.~\ref{fig:classical vortex}. Solutions of this form admit a classical energy which is logarithmically divergent with the size of the box \cite{VortexMass},
\begin{equation}
    E(L) \sim \pi \upsilon^2 \log{\frac{L}{L_0}} \ .
\end{equation}
The divergence of the vortex energy in the infinite volume limit may be interpreted as arising from a cloud of massless Goldstone bosons which surround the vortex.

While topological excitations such as vortices are typically studied either classically or semi-classically at the quantum level, it is interesting to study their fully non-perturbative quantum dynamics. The partition function for the $(2+1)$-d $\U(1)$ lattice quantum field theory is given in the Villain formulation by
\begin{equation}
    \label{eq:villain partition function}
    Z = \pqty{\prod_x \frac{1}{2\pi} \int d\varphi_x} \prod_{\langle x y \rangle} \sum_{n_{\langle x y \rangle} \in \Z} \exp{\bqty{-\frac{1}{2g^2} \pqty{\varphi_x -\varphi_y + 2\pi n_{\langle x y \rangle}}^2}} \ ,
\end{equation}
where $\varphi_x \in [0,2\pi)$ are $\U(1)$ variables associated to each site $x$ of the lattice and $n_{\expval{xy}}$ are integer Villain variables associated with each lattice link $\expval{xy}$. The lattice theory admits a non-trivial continuum limit as one approaches the Wilson-Fisher fixed point. The partition function eq.~\eqref{eq:villain partition function} is exactly dual to the partition function of an integer-valued gauge theory,
\begin{equation}
    \label{eq:integer gauge theory}
    Z =\prod_{\langle xy \rangle} \sum_{A_{\expval{x y}} \in \Z} \exp{\pqty{-\frac{g^2}{2} \sum_{\square} \pqty{dA_\square}^2}} \ ,
\end{equation}
where the product is over all links in the dual lattice and $dA_\square = A_1 + A_2 - A_3 - A_4$ for the four links making up each plaquette $\square$ in the dual lattice. The gauge field $A \in \Z$ is integer-valued. Apart from the fact that the dualized partition function eq.~\eqref{eq:integer gauge theory} for the Villain action eq.~\eqref{eq:villain partition function} is particularly simple, there is nothing special about this choice, and in fact, with appropriate modifications, the same construction also goes through starting from the standard action \cite{VortexMass}.
\begin{wrapfigure}{r}{0pt}
    \centering
    \begin{tikzpicture}
        \draw (0,0) -- (4,0) -- (4,4) -- (0,4) -- cycle;
        \draw (1.5,4) arc (180:275:2.3cm);
        \node at (3,3) {Higgs};
        \node at (1.5,1.2) {Coulomb};
        \node at (-0.5,2.0) {$\kappa$};
        \node at (2.0,-0.5) {$g^2$};
        \node at (-0.25,4.0) {$\infty$};
        \node at (4.0,-0.25) {$\infty$};
        \node at (-0.25,-0.25) {$0$};
    \end{tikzpicture} 
    \caption{Phase diagram of 3D scalar QED.}
    \label{fig:scalar qed phase diagram}
    \vspace{-3em}
\end{wrapfigure}
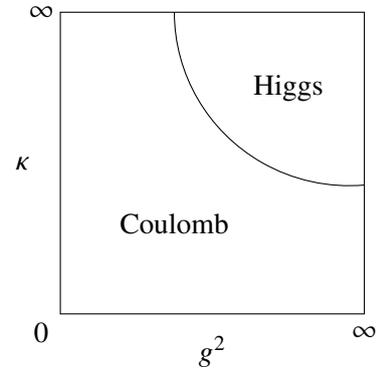
The integer-valued gauge theory admits an interpretation as the $\kappa \to \infty$ limit of non-compact scalar QED \cite{FM} with action 
\begin{equation}
    S[\tilde{A},\phi] = \pqty{\frac{g}{2\pi}}^2 \sum_{\square} \frac12 \pqty{d\tilde{A}_\square}^2 - \kappa \sum_{\langle x y \rangle} \Re \pqty{\Phi_x e^{i \tilde{A}_{\langle x y \rangle}} \Phi_y^*} \ ,
\end{equation}
where $\tilde{A} \in \R$ and $\Phi \in \U(1)$. The coefficient $\frac{1}{e^2} \equiv \pqty{\frac{g}{2\pi}}^2$ of the kinetic term for the gauge field in scalar QED is related to the original coupling via the Dirac quantization condition $ge=2\pi$. In the unitary gauge where $\Phi_x \equiv 1$, taking the limit $\kappa \to \infty$, the field $\tilde{A}$ becomes equal to $2\pi$ times an integer, $\tilde{A}=2\pi A$, and the partition function reduces to the integer-valued gauge theory eq.~\eqref{eq:integer gauge theory}. The phase diagram of scalar QED is summarised in Fig.~\ref{fig:scalar qed phase diagram}. In particular, in the case $\kappa \to \infty$, for small $g^2$ the integer-valued gauge theory is found in a Coulomb phase where the photon is massless. In the original model, the Coulomb phase is mapped to the broken phase, and the photon is dual to the massless Goldstone boson. The vortex in the original theory is dual to the charged particle in scalar QED. On the other hand, for large $g^2$ the integer gauge theory is in a Higgs phase, which is dual to the unbroken phase of the original $\U(1)$ scalar field theory. In this phase the vortices are condensed. The charged particle in scalar QED is given by the gauge-invariant operator of the scalar field together with its non-local Coulomb cloud,
\begin{equation}
    \label{eq:vortex operator}
    \Phi^C_x = \exp{\pqty{i \Delta^{-1} \delta A}} \Phi_x \ ,
\end{equation}
where $\delta$ is the spatial divergence and $\Delta$ the (spatial) Laplacian. It has been rigorously shown that the vortex operator eq.~\eqref{eq:vortex operator} gives rise to topological sectors \cite{FM}. It is worth remarking that charged particles are really \textit{infraparticles} \cite{Buchholz} which are especially sensitive to the boundary conditions.
\begin{figure}
    \centering
    \includegraphics[width=11cm, trim=0cm 0cm 0cm 0cm]{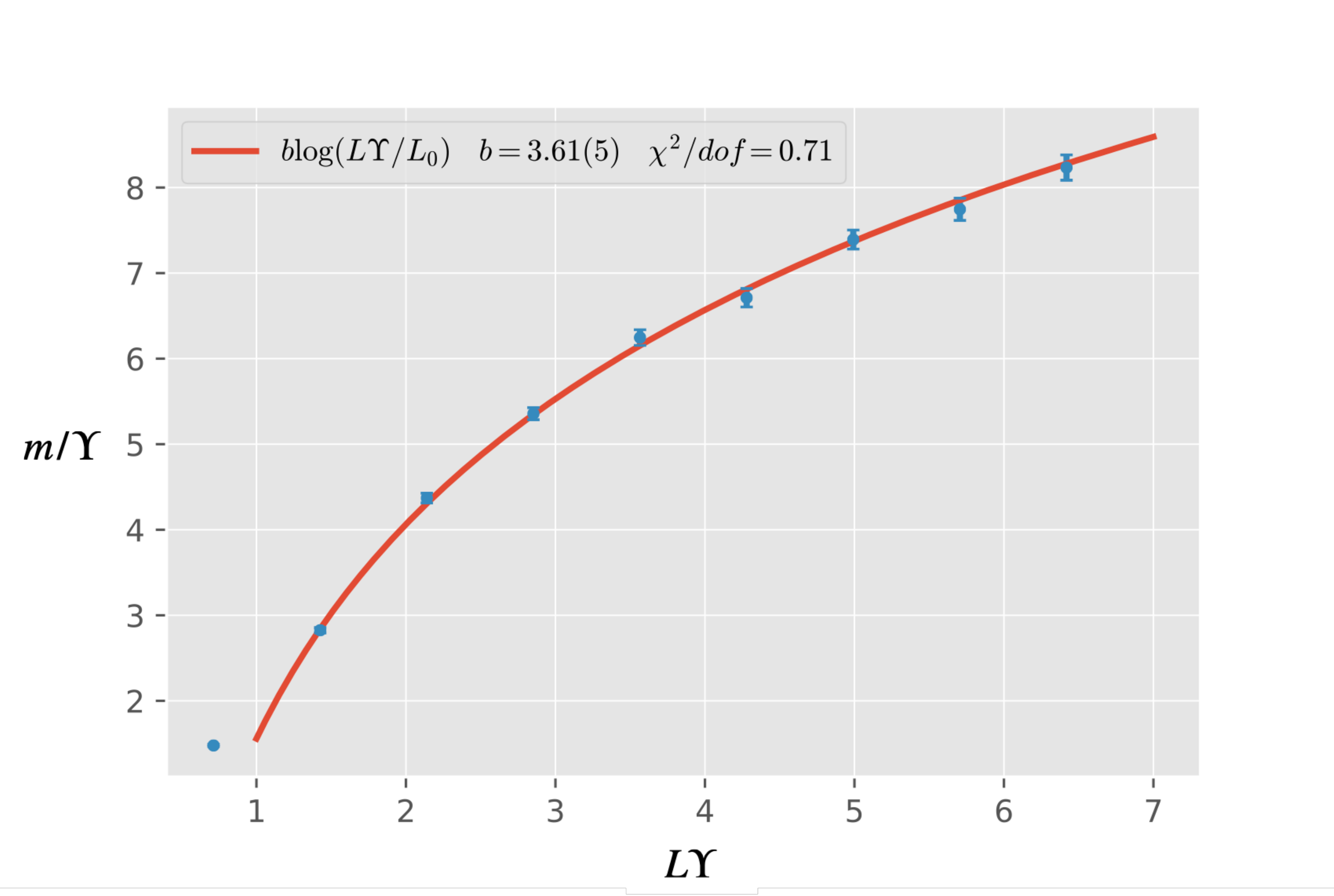}
    \caption{Scaling of the vortex mass in the continuum limit as a function of physical volume, taken from \cite{VortexMass}. Here $\Upsilon$ is the spin stiffness, which is used to set the scale for both $m$ and $L$. The logarithm can be seen to provide an excellent fit for sufficiently large $L$.}
    \label{fig:mass scaling}
\end{figure}
Since in the dual theory the vortex is a charged particle, the simulation of a system with net vortex number cannot be performed using periodic boundary conditions, as explained in the introduction, but one can use $C$-periodic boundary conditions instead. Numerical simulations of a single vortex using $C$-periodic boundary conditions have been performed in \cite{VortexMass}. The mass of the vortex in the broken phase of the $\U(1)$ scalar field theory was measured; one can also measure the vortex condensate in the unbroken phase. Results for the vortex mass in the continuum limit taken at finite physical volume are shown in Fig.~\ref{fig:mass scaling}. The mass is logarithmically divergent with the box size, a result which is qualitatively the same as in the classical theory. We would like to emphasize that there is nothing special with the Villain action eq.~\eqref{eq:villain partition function} and in fact, as an important check of universality, the same results were also obtained starting from the standard action \cite{VortexMass}. It is important to note that, while lacking a confining string, the logarithmic potential is still confining. As we have seen, due to the the logarithmic divergence of their energy, states with non-zero vortex number have infinite energy in the infinite volume. As such, they are absent in the infinite volume theory, which contains only \say{confined} vortex-antivortex pairs with overall zero vortex number. 

It is worth noting that the integer gauge theory has also been studied in four space-time dimensions \cite{PolleyWiese, Jersak}. In that case, it is the dual of four-dimensional pure $\U(1)$ gauge theory and the operator eq.~\eqref{eq:vortex operator} is now the monopole operator, which has been used to obtain the monopole mass and condensate.

\section{Single "Quarks" as Non-Abelian Infraparticles}\label{sec:quarks}

\begin{figure}
    \centering
    \includegraphics[clip, width=0.98\textwidth, trim=15cm 16cm 15cm 16cm]{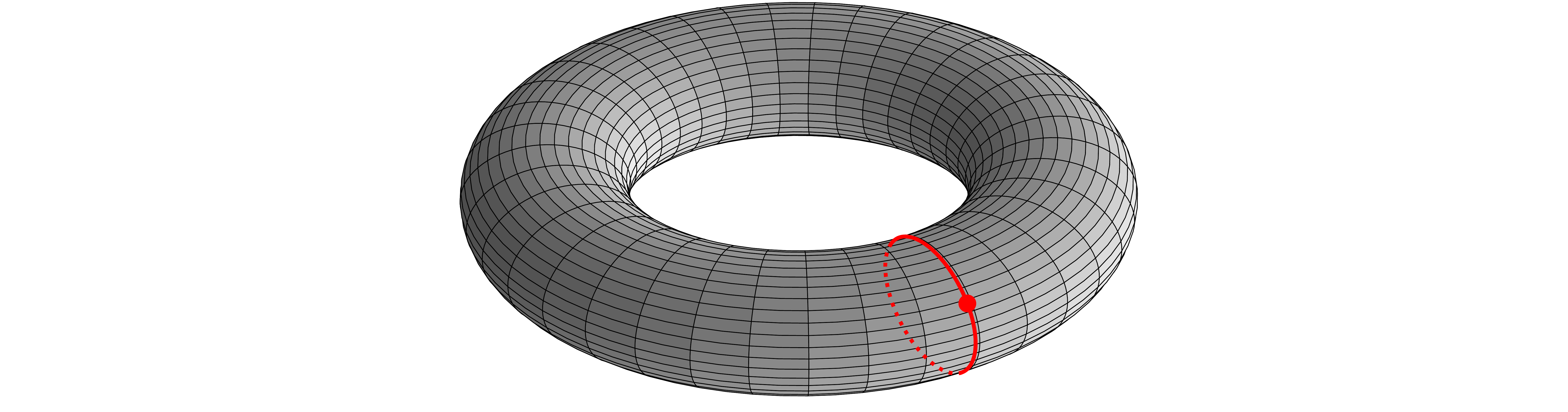}
    \caption{A torus together with a flux string wrapped around a non-contractible loop (thick red line). The dot represents the fermion.}
    \label{fig:flux string}
\end{figure}

As is well known, since QCD is a confining theory it is not possible to observe free individual quarks, which are always confined inside hadrons. This property manifests itself in a pure gauge theory with static external quarks as a linearly rising potential for a quark-antiquark pair with respect to their distance. In a gauge theory with dynamical fermions such as QCD, however, the manifestation of confinement is more subtle \cite{FredMarcu}. In principle, confinement should be thought of as a dynamical property of QCD, rather than simply a statement about the possible gauge-invariant states. Therefore, it is conceivable that in a finite volume it may be possible to construct a gauge-invariant single-quark state. Since it would be a charged state, under both $\U(1)_{\mathrm{em}}$ and $\SU(N)_c$, this is not possible in a periodic volume, but it may be possible in a $C$-periodic volume. Confinement would then manifest itself as the dynamical property of such a state having an energy which is linearly divergent with the size of the box.

\begin{wrapfigure}{R}{0pt}
    \centering
    \vspace{-20pt}
    \includegraphics{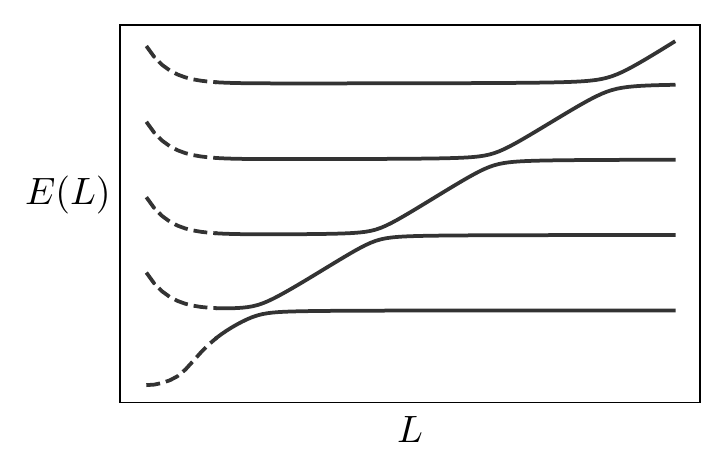}
    \caption{Schematic view of finite-volume energies $E(L)$ indicating the presence of a quark state with linearly rising energy mixing with baryon states. In a realistic simulation the various states are unlikely to be so clearly separated.}
    \label{fig:quark state level crossings}
    \vspace{-10pt}
\end{wrapfigure}

For $\SU(N)$ gauge theory in a $C$-periodic volume, we have seen above that baryon number is partially broken to fermion number $\mathbb{Z}(2)_F$. For even $N$, all colorless states are bosonic, and they can therefore be distinguished from a fermionic quark state. On the other hand, for odd $N$, both quarks and baryons are fermionic and it is then possible for an operator which creates a quark state to have non-zero overlap with baryon states. As such, if a gauge-invariant quark operator can be constructed for even $N$, it would provide a clean channel to investigate dynamical properties of confinement, whereas for odd $N$ more complicated finite-volume studies would be necessary to distinguish properties of a quark state from ordinary baryon states, as depicted in Fig.~\ref{fig:quark state level crossings}.

Explicitly, a gauge-invariant single-quark creation operator is straightforward to define for odd $N$ using a composite operator consisting of an ordinary fermion operator together with a flux string winding around a non-contractible $C$-periodic spatial loop $\mathcal{C}$, as shown in Fig.~\ref{fig:flux string}. This is gauge-invariant because spatial Polyakov loops winding around $C$-periodic boundaries transform unusually,
\begin{equation}
    U_{\mathcal{C}}(\vec x) = \mathcal{P} \exp{\pqty{\int_0^{L_i} dy_i\, G_i (\vec x + y_i \vec{e}_i)}} \quad \implies \quad U'_{\mathcal{C}}(\vec x) =\Omega(\vec{x}) U_{\mathcal{C}}(\vec x) \Omega(\vec{x})^T \ .
\end{equation}
For a quark in the fundamental representation of $\SU(3)$, a gauge-invariant operator is then given by
\begin{equation}
    \label{eq:quark fundamental SU3}
    \Psi_{\mathcal{C}}(\vec x) = \epsilon_{abc}\psi^a U_{\mathcal{C}}^{bc}(\vec x).
\end{equation}
Similar operators can be defined for larger odd $N$ using additional Polyakov loops.

The inclusion of Polyakov lines in the operators used to create quark states for odd $N$ suggests that these states may still be distinguished from ordinary baryon states by their center charge rather than baryon number. Unfortunately, as argued above, $C$-periodic boundaries also break the $\mathbb{Z}(N)$ center symmetry completely for odd $N$. On physical grounds, this is not a surprising outcome: each Polyakov line can be broken by the insertion of a quark $-$ anti-quark pair without modifying the quantum numbers of the operator and these pairs can be separated and taken around the $C$-periodic boundary in a smooth process to yield a fully local baryonic operator; as such, the baryon and quark states must necessarily appear with identical quantum numbers.

This construction is not possible for even $N$, because it is impossible to saturate the $N$ indices of the $\epsilon$-tensor with one fermionic operator and any number of Polyakov loops. It is natural to then consider more general fermionic, gauge-invariant operators; however, a parity argument in the Hamiltonian picture quickly shows that no such construction is possible when the quark lives in a representation of $\SU(N)$ with odd $N$-ality,\footnote{The $N$-ality of an $\SU(N)$ representation is given by the number of boxes in the associated Young tableau.} which includes the fundamental representation. This argument, however, does not apply to representations with even center charge. For example, in $\SU(4)$ gauge theory, quark states in even $N$-ality representations can be readily constructed, including
\begin{align}
    \mathbf{6}: \quad &\Xi_{\mathcal{C}}(\vec x) =  \epsilon_{abcd} \xi^{ab}(\vec{x})U_{\mathcal{C}}^{cd}(\vec x) \ , \\
    \mathbf{10}: \quad & E_{\mathcal{C}}(\vec x) =  \eta^{ab}(\vec{x})U_{\mathcal{C}}^{ab}(\vec x) \ ,
\end{align}
where $\xi$ and $\eta$ are fermion fields respectively symmetric and antisymmetric in their color indices. As a result of the above arguments, it is possible to construct single-\say{quark} states in appropriate representations of $\SU(N)$ for even $N$, however interestingly not in the fundamental representation. We expect these states to have an energy which is linearly divergent with the size of the finite volume. Rather peculiarly, fundamental-representation quark-like states may be constructed and investigated only for odd $N$, but in such cases they are not distinguished from ordinary baryon states by any quantum number. Though it would require a complicated finite-volume study, it would nevertheless be interesting to investigate the spectrum in the fermionic sector for $\SU(N)$ gauge theories with odd $N$ to determine whether a quark-like state in the spectrum can be isolated.

\section{Conclusions}

Since their introduction in \cite{PolleyWiese, KronfeldWiese}, $C$-periodic and $G$-periodic boundary conditions have found several applications, for example in the numerical simulation of QED corrections to QCD \cite{Patella} and in the computation of the kaon decay amplitude on the lattice \cite{Christ}.

In this work, we focused on the construction of confined charged particles in a finite $C$-periodic volume. We have seen that vortices in the $(2+1)$-d $\U(1)$ scalar field theory can be quantized fully non-perturbatively. The construction exploits a duality transformation, whereby the vortices are identified with charged particles in the dual theory, surrounded by a cloud of massless Goldstone bosons (which are dual to photons). Because of their infraparticle nature, they are especially sensitive to the boundary conditions. With $C$-periodic boundary conditions in the dual theory, the vortices have a finite, well-defined mass in a finite volume, and their mass diverges logarithmically with the size of the finite box.

For $\SU(N)$ gauge theories, $C$-periodic boundary conditions break baryon number and center symmetry completely for odd $N$ and partially for even $N$. As a result, in $C$-periodic QCD, and more generally in $\SU(N)$ gauge theories with odd $N$, the single-quark states mix with baryon states. On the other hand, for all even $N \geq 4$ it is possible to construct stable single-\say{quark} states as non-Abelian infraparticles. Their energy, which we expect to diverge linearly with increasing volume, defines a physical \say{quark} mass that runs with the volume. One can think of the quark state as a kind of fermionic string, and it would be interesting to study it in the spirit of effective string theory \cite{LuscherWeisz}.

\acknowledgments

GK, AM and UJW acknowledge funding from the Schweizerischer Nationalfonds (grant agreement number 200020\_200424).


\begin{thebibliography}{99}

\bibitem{PolleyWiese}
L. Polley and U.-J. Wiese,
\emph{Monopole condensate and monopole mass in U(1) lattice gauge theory},
\emph{Nucl. Phys. B} {\bf 356} (1991) 629.

\bibitem{KronfeldWiese}
A. Kronfeld and U.-J. Wiese,
\emph{SU(N) gauge theories with C-periodic boundary conditions (I). Topological structure},
\emph{Nucl. Phys. B} {\bf 357} (1991) 521.

\bibitem{WieseCG}
U.-J. Wiese,
\emph{C-periodic and G-periodic QCD at finite temperature},
\emph{Nucl. Phys. B} {\bf 375} (1992) 45.

\bibitem{VortexMass}
M. Hornung, J. C. Pinto Barros and U.-J. Wiese,
\emph{Mass and Charge of the Quantum Vortex in the (2+1)-d O(2) Scalar Field Theory}.
[{\tt arXiv:2106.16191}]

\bibitem{FM}
J. Fr\"ohlich and P. A. Marchetti,
\emph{Soliton quantization in lattice field theories},
\emph{Commun. Math. Phys.} {\bf 112} (1987) 343.

\bibitem{Buchholz}
D. Buchholz,
\emph{Gauss' law and the infraparticle problem},
\emph{Phys. Lett. B} {\bf 174} 3 (1986) 331.

\bibitem{Jersak}
J. Jersák, T. Neuhaus, and H. Pfeiffer,
\emph{Scaling analysis of the magnetic monopole mass and condensate in the pure U(1) lattice gauge theory},
\emph{Phys. Rev. D} {\bf 60} (1999) 054502.
[{\tt arXiv:hep-lat/9903034}]

\bibitem{FredMarcu}
K. Fredenhagen and M. Marcu,
\emph{Confinement criterion for QCD with dynamical quarks},
\emph{Phys. Rev. Lett.} {\bf 56} (1986) 223.

\bibitem{Patella}
B. Lucini, A. Patella, A. Ramos, and N. Tantalo,
\emph{Charged hadrons in local finite-volume QED+QCD with C\textsuperscript{*} boundary conditions},
\emph{JHEP} {\bf 02} (2016) 076.
[{\tt arXiv:1509.01636}]

\bibitem{Christ}
Z. Bai, T. Blum, P.A. Boyle, N.H. Christ, J. Frison, N. Garron, T. Izubuchi, C. Jung, C. Kelly, C. Lehner, R.D. Mawhinney, C.T. Sachrajda, A. Soni, and D. Zhang (RBC and UKQCD Collaborations),
\emph{Standard Model Prediction for Direct CP Violation in $K \to \pi \pi$ Decay},
\emph{Phys. Rev. Lett.} {\bf 115} (2015) 212001.
[{\tt arXiv:1505.07863}]

\bibitem{LuscherWeisz}
M. L\"uscher and P. Weisz,
\emph{String excitation energies in SU(N) gauge theories beyond the free-string approximation},
\emph{JHEP} {\bf 0407} (2004) 014.
[{\tt arXiv:hep-th/0406205}]

\end{thebibliography}
\end{document}